\documentclass[conference]{IEEEtran}
\IEEEoverridecommandlockouts
\usepackage{cite}
\usepackage{amssymb,amsfonts}
\usepackage{amsmath}
\usepackage{algorithm}
\usepackage{algpseudocode}
\usepackage{graphicx}
\usepackage{textcomp}
\usepackage{xcolor}
\usepackage{fancyhdr}
\usepackage{algpseudocode}
\usepackage{subfigure}
\def\BibTeX{{\rm B\kern-.05em{\sc i\kern-.025em b}\kern-.08em
    T\kern-.1667em\lower.7ex\hbox{E}\kern-.125emX}}

\newboolean{showcomments}
\setboolean{showcomments}{true}
\ifthenelse{\boolean{showcomments}}
{
}

\newcommand*{\affmark}[1][*]{\textsuperscript{#1}}

\begin{document}

\title{Hierarchical Reinforcement Learning Based Traffic Steering in Multi-RAT 5G Deployments}

\author{\IEEEauthorblockN{Md~Arafat~Habib\affmark[1], Hao~Zhou\affmark[1], Pedro~Enrique~Iturria-Rivera\affmark[1], Medhat Elsayed\affmark[2], Majid Bavand\affmark[2], \\Raimundas Gaigalas\affmark[2], Yigit Ozcan\affmark[2] and  Melike Erol-Kantarci\affmark[1], \IEEEmembership{Senior Member,~IEEE}}
\IEEEauthorblockA{\affmark[1]\textit{School of Electrical Engineering and Computer Science, University of Ottawa, Ottawa, Canada}}  \affmark[2]\textit{Ericsson Inc., Ottawa, Canada}\\
Emails:\{mhabi050, hzhou098, pitur008, melike.erolkantarci\}@uottawa.ca, \\\{medhat.elsayed, majid.bavand, raimundas.gaigalas, yigit.ozcan\}@ericsson.com \vspace{-1em}}

\maketitle

\thispagestyle{fancy}   
\fancyhead{}                
\lhead{Accepted by 2023 IEEE International Conference on Communications (ICC) , \copyright2023 IEEE}
\cfoot{}
\renewcommand{\headrulewidth}{0pt} 

\begin{abstract}
In 5G non-standalone mode, an intelligent traffic steering mechanism can vastly aid in ensuring smooth user experience by selecting the best radio access technology (RAT) from a multi-RAT environment for a specific traffic flow. In this paper, we propose a novel load-aware traffic steering algorithm based on hierarchical reinforcement learning (HRL) while satisfying diverse QoS requirements of different traffic types. HRL can significantly increase system performance using a bi-level architecture having a meta-controller and a controller. In our proposed method, the meta-controller provides an appropriate threshold for load balancing, while the controller performs traffic admission to an appropriate RAT in the lower level. Simulation results show that HRL outperforms a Deep Q-Learning (DQN) and a threshold-based heuristic baseline with 8.49\%, 12.52\% higher average system throughput and 27.74\%, 39.13\% lower network delay, respectively. 
\end{abstract}


\section{Introduction}
Managing user traffic in non-standalone (NSA) fifth generation new radio (5G NR) is challenging since the traffic can be either directed to 5G or long term evolution (LTE) network. In NSA, user equipments (UE) access the multiple radio access technologies (multi-RAT) using dual connectivity (DC) \cite{16,20}. Each type of RAT has different abilities to provide service to the UE with diverse quality of service (QoS) requirements. However, if the traffic is always steered to the same base station with a certain RAT that can best serve the QoS requirements, this may result in unbalanced load distribution. This will eventually cause high delay in the network leading to packet drops, causing a negative impact on the average system throughput. If traffic with high load arrives and data packets are aggregated into a flow, they cannot be segregated again \cite{28}. Therefore, if a packet is forwarded to a congested queue, it will suffer from long waiting time until the queue is emptied. Considering all these issues, it becomes important to develop a load-aware robust traffic steering scheme. 

Throughput-hungry applications that emerged after  5G deployments, have reached an unprecedented level \cite{25}. These applications require stringent fulfillment of QoS demands along with flexible and intelligent network management entities. Furthermore, architectural reformation introduced in open radio access network (O-RAN) \cite{29,34,35} can facilitate RAN with openness and required intelligence. The radio controller in an O-RAN architecture is divided into two parts, near-real-time RAN intelligent controller (near-RT-RIC) and non-real-time RAN intelligent controller (non-RT-RIC). The non-RT-RIC is in the top of the hierarchy that serves as a software platform for the designed rApp for high level RAN optimization. It has visibility into network information, provides artificial intelligence (AI)-enabled feeds and recommendations to near-RT-RIC. Near-RT-RIC in the lower level enables control and optimization of RAN elements. Programmable and highly modular structure of future disaggregated RANs is quite suitable for developing advanced AI-based modules to perform network optimization via robust traffic steering schemes.

A machine learning (ML)-enabled traffic steering scheme can be a great tool to optimize network performance in a multi-RAT environment. Considering this fact, attempts have been made to design traffic steering schemes for 5G using ML, specially using reinforcement learning (RL) \cite{2}. RL algorithms can provide us with the ease of avoiding any dedicated optimization model since we can transform  optimization problems into Markov decision processes (MDPs). Furthermore, compared to conventional RL algorithms, hierarchical reinforcement learning (HRL) can provide better exploration efficiency via meta-controller and controller instead of using a standalone agent \cite{25}. In particular, the bi-level architecture of O-RAN having near and non-RT-RIC makes it a suitable candidate to embed the meta-controller and controller in the O-RAN hierarchy as rApps and xApps. Therefore, different from the previous works, we propose an HRL-based traffic steering algorithm that is applicable for O-RAN architecture. 

In this paper, we intend to maintain QoS requirements of all the traffic types simultaneously by proposing an HRL-based traffic steering scheme that at the same time is able to perform threshold-based load balancing in a disaggregated RAN environment. The threshold is associated with the queue length of each RAT which is provided by the meta-controller and the controller in the lower level is responsible for RAT specific traffic steering. We compare the performance of the proposed method with a deep reinforcement learning (DRL)-based baseline namely deep Q-learning (DQN) and a threshold-based heuristic baseline. The proposed scheme gains as high as 8.49\%  and 12.52\% improved average system throughput and 27.74\% and 39.13\% lower network delay compared to the DQN and threshold-based heuristic baseline algorithm, respectively.    

We organize the remaining parts of the paper as follows: Section \ref{s2} and \ref{s3} discuss the related works to our research conducted in this paper, and the system model along with problem formulation, respectively. The proposed HRL-based traffic steering algorithm is covered in Section \ref{s4}. Performance comparison of the proposed method along with the baseline algorithms is presented in Section \ref{s5}. Finally, conclusions are presented in Section \ref{s6}.

\section{Related work}
\label{s2}
Network optimization has been conducted via traffic steering schemes in the literature after the emergence of 4G/LTE networks. 5G deployments can be benefited highly by an efficient traffic steering scheme since it can vastly help to deal with the increased number of users with DC, multiple traffic types, and dense establishment of small cells. A threshold-based traffic steering scheme is proposed in \cite{1} that performs network optimization based on channel condition, load level at each RAT, and service type. Dryjanski et al. propose a traffic steering use case for ORAN with predefined policies in which xApps are designed for spectrum management, cell assignment and resource allocation \cite{23}.

In recent years, RL algorithms have been used several times in the literature to develop traffic steering schemes for 5G multi-RAT environment in O-RAN. Fetemeh at al. propose an intelligent traffic steering scheme for O-RAN to handle unknown traffic demand using recurrent neural network \cite{32}. Cao et al. develop a federated learning-based scheme for O-RAN, in which each UE acts as an agent to make network access decisions independently \cite{24}. An O-RAN based RAT allocation environment that enables RL agents to train their DQN models for steering their traffic between RATs is proposed in \cite {26}. In comparison to the existing literature, contributions of our proposal lie in developing an automated traffic steering mechanism specific to each RAT that can maintain QoS requirements of different traffic types. Furthermore, it can provide interruption less network connectivity ensuring smooth user experience via threshold based load balancing in 5G NSA mode using HRL. 

\section{System Model and Problem Formulation}
\label{s3}
\subsection{System Model}
\begin{figure}[!t]
\centerline{\includegraphics[width=0.7\linewidth]{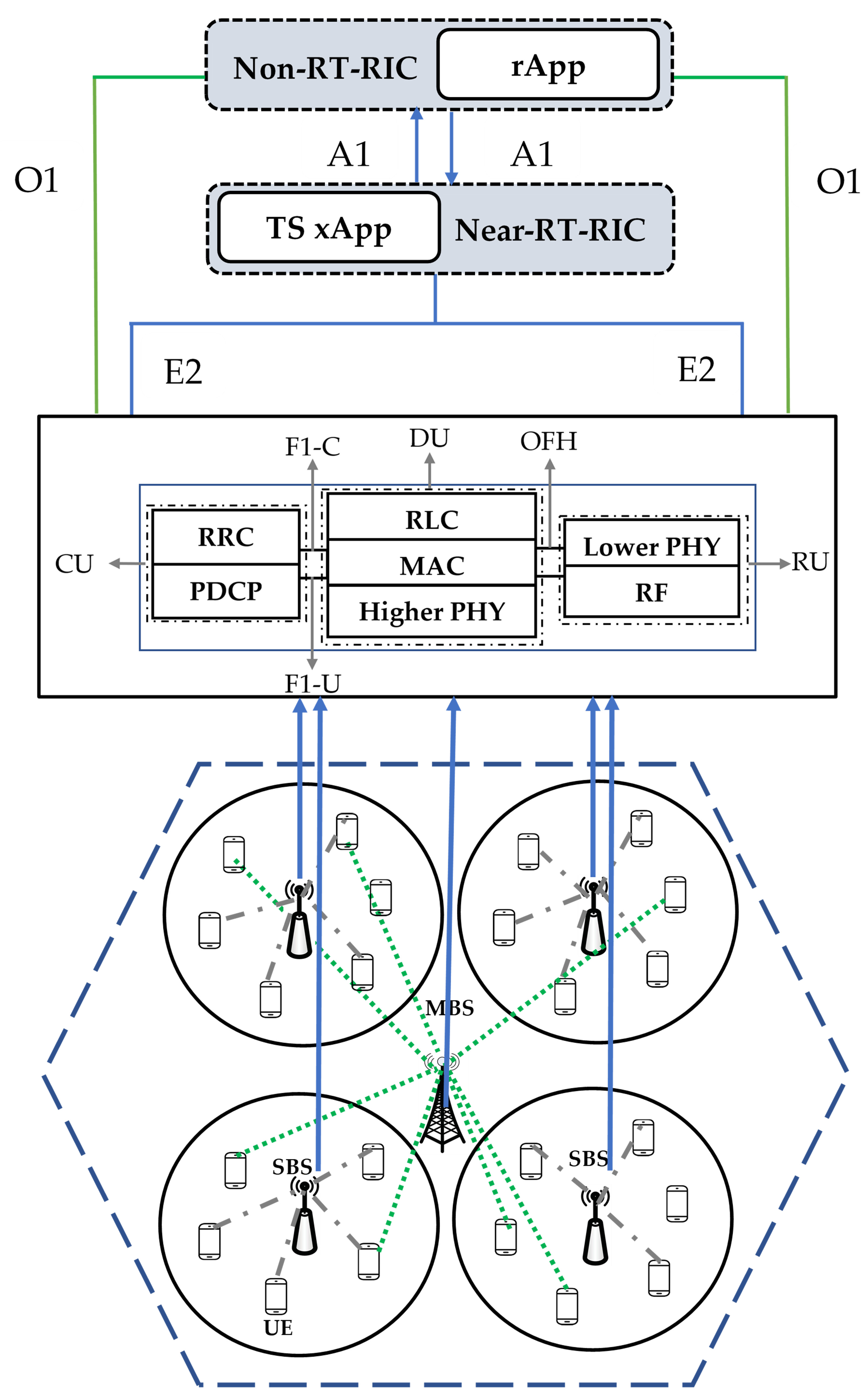}}
\caption{5G NSA deployment with macro and small cells controlled via intelligent controllers.}
\label{fig1}
\vspace{-1em}
\end{figure}

In this work, we consider a multi-RAT network where multiple users are connected with 5G and LTE RATs via DC. There are small cells having 5G NR base stations (BSs) that can serve applications requiring high throughput and low latency. Small cells are within the range of a macro-cell that is facilitated by one LTE BS (eNB). There are total $\Phi$ flows in the network and each UE in the network has a traffic flow $\phi$, that can be either steered to eNB or gNB based on the decision of the HRL agent. The considered wireless system for 5G NSA mode is presented in Fig. \ref{fig1}.

There are two control loops in the system. First one is the non-RT control loop which has a latency larger than 1s. This is where policies are set, and RAN analytics are gathered. The second control loop operates in larger than 10 ms and less than 1s time frame. In this time frame, our traffic steering xApp operates and produces actions to perform flow admission to a RAT.  

The total downlink bandwidth, $B$ (in MHz) is divided into $X_{RB}$ resource blocks. Each RBG, $\psi$ is assigned some transmission power $\rho_{\psi,b}$ by a BS, $b$. The link capacity between UE, $u$ and BS, $b$ can be formulated as below: 

\begin{equation}    \xi_{u,b}=\sum_{\psi=1}^{\Psi}\omega_\psi\log_2\left(1+\frac{\rho_{\psi,b}\zeta_{\psi,u}g_{\psi,u,b}}{\omega_\psi X_0+\sum_{\mu\in B}\rho_{\psi,\mu}\zeta_{\psi,u,\mu}g_{\psi,u,\mu}}\right),
\end{equation}
where $\omega_\psi$ is the bandwidth of the $\psi$, $\rho_{\psi,b}$ is the transmit power of the BS, $b$ on $\psi$, $g_{\psi,u,b}$ is the channel co-efficient and $\zeta_{\psi,u,b}$ is the RBG's allocation indicator of the link $(\psi,u,b)$. $X_0$ is the additive white Gaussian noise single-sided power spectral density. $\rho_{\psi,\mu}$ is the transmit power of the interfering BS, $g_{\psi,u,\mu}$ is the channel co-efficient, and $\zeta_{\psi,u,\mu}$ is the allocation indicator of link $(\psi,u,\mu)$. The link capacity should not be exceeded as traffic flows pass through a link in the system
\begin{equation}
    \sum_{\phi\in \Phi}\delta_{\phi}v_{u,b}^{\phi}\leqslant \xi_{u,b}\quad \forall(u,b)\in L,
\end{equation}
where capacity demand of a flow is represented using $\delta_\phi$, $v_{u,b}^{\phi}$ is a binary variable which is `1' given that the link $(u,b)$ has been used from $u$ to BS $b$. It is `0' otherwise. $L$ is the set of links and $\Phi$ is the set of all the traffic flows.  

The proposed system model considers delay as the combination of transmission and queuing delay which is as follows $D_{k,b}=D_{k,b}^{T}+D_{k,b}^Q$,
where $D_{k,b}^{T}$ is the transmission delay experienced for a specific traffic type $k$, and $D_{k,b}^Q$ is the queuing delay that took place for a certain traffic type $k$ at BS $b$ for a user $u$. 

\subsection{ Problem Formulation and QoS Requirements}

To conduct traffic steering for different traffic types having variant QoS requirements for delay and throughput, we define two parameters. First one is the delay parameter which is calculated as the ratio of the defined QoS requirement for delay ($D_{QoS}$) and the actual delay ($D_{k,b}$) experienced in the system for a specific traffic type ($k$). It is formulated as follows: 
\begin{equation}
    \varpi_{k,b}^D=\frac{D_{QoS}}{D_{k,b}}.
\end{equation}

Similarly, we get the throughput parameter as the ratio of the throughput achieved by the system ($T_{k,b}$) running our algorithm and the minimum throughput required ($T_{QoS}$)
\begin{equation}
    \varpi_{k,b}^T=\frac{T_{k,b}}{T_{QoS}},
\end{equation}

The goal of the proposed scheme is to improve the system performance in terms of delay and throughput. To represent such goal, a new variable is initiated. The variable combines the delay and throughput parameters that were presented in eq. (3) and (4). It is as follows:
\begin{equation}
    P=c_1(\varpi_{k,b}^D)+c_2(\varpi_{k,b}^T)-H,
\end{equation}
where $c_1$ and $c_2$ are the weight factors and $H$ is the handover penalty since excessive handover in the system would affect the system throughput. The network optimization problem that we want to address in this work is associated with this variable $P$, and is as follows:  

\begin{equation}
     \begin{split}
      max\sum_{u\in U}\sum_{\phi\in K}\sum_{b\in B}P_{u,\phi,b}, \quad \quad \quad \quad \\
     s.t. \sum_{(u,b)\in L}\beta^{\phi_k}\geqslant \beta^\phi \quad \forall \phi \in \Phi, \quad \quad\\
     \sum_{(u,b)\in L}D(u,b)v_{u,b}^\phi\leqslant D^\phi \quad \forall \phi \in \Phi, \quad
     \end{split}
\end{equation}
where $\beta^{\phi_k}$ is the required bitrate for a particular type of traffic $k$, and $\beta^\phi$ is the available bitrate. Also, $D^\phi$ represents the latency demand of flow $\phi\in \Phi$ and $D(u,b)$ is the latency of link $(u,b)$.  

On one hand, steering the user traffic to the RAT that can best serve the QoS demands of that specific traffic type can significantly increase network performance. However, for the long term performance, it is necessary that we consider the high load imposed to the BS when the traffic load increases vastly. Therefore, to maximize the total objective, we have to provide an intelligent mechanism that can perform load balancing to satisfy the desired performance. Considering that, we introduce the HRL algorithm that can perform threshold-based load balancing in a dynamic manner.  

\section{Proposed HRL-based Traffic Steering Scheme}
\label{s4}
In this section, we describe the proposed HRL-based traffic steering scheme. First, terminologies and notations related to HRL are discussed in brief and MDPs are defined. Next, we present how Q-values are updated and goals are selected. Finally, the load-aware HRL-based traffic steering algorithm is presented.      

\subsection{Hierarchical Reinforcement Learning}

In typical RL, the problem is defined as an MDP $<S,A,T,R>$, where $S$ is the set of states, $A$ is the set of actions, $T$ is the transition probability ($T:S\times A\times S$), and $R$ is the reward function. A standalone agent interacts with the environment to maximize the reward \cite{30}. 

Compared with the traditional RL, the agent consists of two controllers, meta-controller and controller in HRL \cite{31}. The MDP for HRL is rewritten as $<S,A,T,R,G>$. $G$ in the tuple indicates a set of goals. Based on the current state $s \in S$, the meta-controller is supposed to produce high level goals $g \in G$ for the controller. Next, these goals are transformed for high-level policies. The controller is responsible for choosing low-level actions $a \in A$ based on the high-level policies and on the process of doing so, receives an intrinsic reward ($r_{in}$). Finally, the meta-controller will get an extrinsic reward ($r_{ex}$) from the environment and provide the controller with a new goal, $g\prime$. HRL can provide with more efficient learning because of the hierarchy introduced in the architecture. By dividing the sub-goals, HRL allows more efficient management of RAN functionalities.

\begin{figure}[!t]
\centerline{\includegraphics[width=0.7\linewidth]{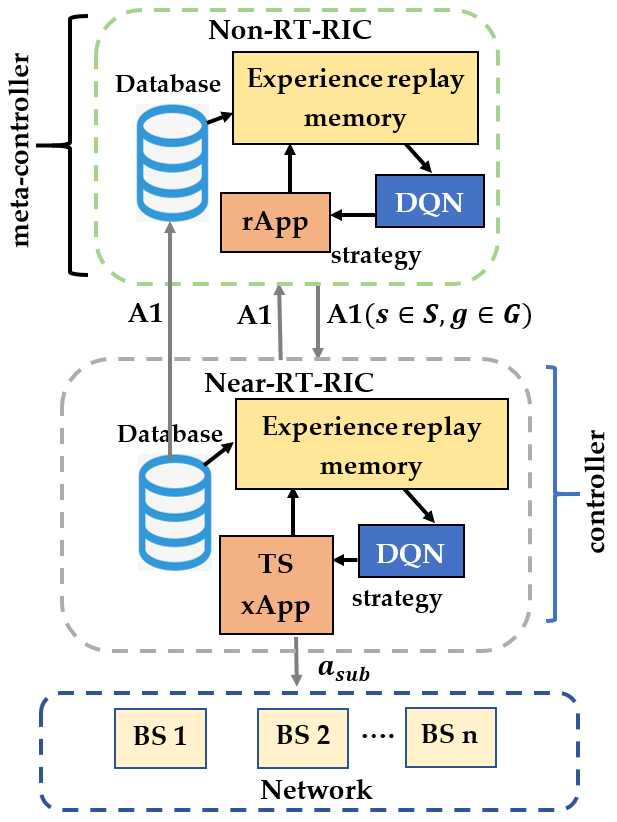}}
\caption{HRL integration with intelligent controllers at different timescales.}
\label{fig2}
\vspace{-1.2em}
\end{figure}

In our proposed hierarchical implementation of intelligence, the HRL-model takes decisions at two time scales. Meta-controller on top level module (could be placed in non-RT-RIC) takes in the state perceived by the agent from the network environment and picks a new load balancing threshold as a goal. On the other hand, the controller which can be embedded in near-RT-RIC uses both the state and the chosen goal to select the actions until the episode is terminated. The models are trained using stochastic gradient descent at different temporal scales to optimize expected future intrinsic reward for the xApp-based controller and extrinsic reward for the rApp-based meta-controller. To summarize the whole process, we present the schematic of the proposed method in Fig. \ref{fig2}. 

To transform the problem formulated in eq. (6) into HRL notations, the following MDP is defined for the meta-controller and controller. 

\begin{itemize}
\item \textbf{State:} There are three elements in the set of states, $s_{con}=\{F_{t},SINR_r,Q_l$\}. Here, $F_{t}$ represents the traffic flow. The second element of the set of states is the SINR measurements to represent the link quality between a BS and UE: $SINR_r$=\{$SINR_{LTE},SINR_{NR}$\}. As for the last element in state space, we use queue length of both LTE and 5G NR RATs to represent load level: $Q_l$=\{$Q_{l(NR)},Q_{l(LTE)}\}$.
\item \textbf{Action:} Flow admission to the different RATs in a multi-RAT environment is considered in the action space which is defined as: \{$A_{L},A_{NR}\}$. Here, flow admission to the LTE RAT is presented by $A_{L}$ and 5G RAT by $A_{NR}$.
\item \textbf{Intrinsic reward:} The intrinsic reward function ($r_{in}$) for the controller is same as eq. (5).
\end{itemize}

The meta-controller is responsible for high level policies for the agent. MDP definition for the meta-controller is stated as follows:

\begin{itemize}
\item \textbf{State:} The states of the meta-controller consists of the traffic type, SINR measurements, and queue length of each type of RAT: $s_{meta}=\{F_{t},SINR_r,Q_l$\}.
\item \textbf{Goal for the controller:} Thresholds associated with the queue length is considered as the goals for the controller. Therefore, $G=\{g_1,g_2,..,g_n\}=\{Th_1,Th_2,...,Th_n\}$. Transmission is differed to another RAT for load balancing based on this threshold. 
\item \textbf{Extrinsic reward:} The meta-controller is responsible for the overall performance of the whole system. Therefore, we have set the extrinsic reward function for the meta-controller as the objective of the problem formulation presented in eq. (6). The following equation is basically the summation of the intrinsic reward over $\tau$ steps. 

\begin{equation}
r_{ex,\tau}= \frac{1}{n}\sum_{\tau=1}^{n} r_{in,\tau} \quad 
\forall(u)\in U,\forall(b)\in B,
\end{equation}

\end{itemize}

\subsection{Q-value Update and Selection of Goals}

In this section of the paper, we present how to update the Q-values of the controller along with the action and goal selection strategies. Q-values of the meta-controller is updated by: 

\begin{equation}
\begin{split}
    Q_{M}^{N}(s_{meta},g_{meta})=Q_{M}^{O}(s_{meta},g_{meta}) +\alpha(r_{ex}+ \\
    \gamma\max_{g}Q_{meta}(s_{meta}^\prime,g,\theta_1)-Q_{M}^{O}(s_{meta},g_{meta},\theta_{1}^\prime)),
\end{split}
\end{equation}
where $s_{meta}^\prime$ is the next state, $\alpha$ is the learning rate, and $\gamma$ is the discount factor. $\theta_1$ and $\theta_{1}^\prime$ are the weights associated with the main network and the target network. The new and old values are represented as $Q_{M}^{N}$ and $Q_{M}^{O}$. This means the accumulated reward is brought by state-goal pair ($s_{meta},g_{meta}$). Next, we use the $\epsilon$-greedy policy for goal selection which can balance the exploration and exploitation of goals so that long term rewards are achieved. 

\begin{equation}
  \pi(s_{meta})=\begin{cases}
    \text{random goal selection}, & \text{$rand \leqslant \epsilon$} \\
    arg \max_{g} Q(s_{meta},g), & \text{$rand > \epsilon$},
  \end{cases}
\end{equation}
where $rand$ is a random number generated between 0 to 1 and $\epsilon$ is less than 1.

We update the Q-values of the controller using:
\begin{equation}
\begin{split}
    Q_{C}^{N}(s_{con},g_{meta},a_{con})=Q_{S}^{O}(s_{con},g_{meta},a_{con})\\ +\alpha(r_{in}+ \gamma\max_{a}Q_{con}(s_{con}^\prime,g_{meta}^\prime,a,\theta_2)-\\Q_{S}^{O}(s_{con},g_{meta},a_{con},\theta_{2}^\prime)) \quad \quad \quad,
\end{split}
\end{equation}
where $s_{con}^\prime$ is the next state and the next goal produced by the meta-controller is $g_{meta}^\prime$. $\theta_2$ and $\theta_{2}^\prime$ are the weights associated with the main network and the target network,respectively. The old and new Q-values for the controller are represented by $Q_{S}^{O}$ and $Q_{S}^{N}$, respectively. Like before, we use the $\epsilon$-greedy policy for controller's action selection. 

\subsection{Baseline Algorithms}

In this work we are using two baselines. First one is the DQN-based traffic steering scheme that uses similar system model and a static load balancing threshold \cite{33}.

To show the performance improvement of the proposed HRL-based traffic steering scheme algorithm, we will compare it with a threshold-based heuristic algorithm \cite{1}. The algorithm utilizes a predefined threshold calculated using load at each station, channel condition, and user service type. The threshold ($Th_t$) is calculated by considering the mean of all the metrics mentioned. A variable $W$ is computed using the same parameters with weight metrics (through summation). Traffic steering decision is taken through the comparison of the $W$ and $Th_t$.

\section{Performance Evaluation}
\label{s5}
\subsection{Simulation setup}
We deploy a MATLAB-based simulation environment that includes one eNB and four gNBs which serve one macro-cell and four small cells. There are 60 users in the simulation environment. There are three types of traffic: Video, Gaming and Voice traffic. Video traffic in our simulation setting had the highest throughput requirement. To test how the proposed traffic steering algorithm performs with high throughput requirement we have set the proportion of the video traffic to be 50\%. Gaming traffic in the system has the most precise delay requirement and we have 30\% proportion of that. Lastly, proportion of the voice traffic is 20\%. QoS requirements of different traffic types have been defined based on 3GPP specifications and specifications presented in \cite{14}. Packet size, $T_{QoS}$, and $D_{QoS}$ are considered to be 30 bytes, 0.1 Mbps, and 100 ms, respectively for the voice traffic. Same parameters are specified to be 250 bytes, 10 Mbps, and 80ms for the video traffic. Lastly, for the gaming traffic, we set packet size, $T_{QoS}$, and $D_{QoS}$ to be 120 bytes, 5Mbps, and 40ms, respectively. 

We consider multi-RAT environment in 5G NSA mode where LTE and 5G NR BSs serve together. An architecture described in \cite{13} has been opted for implementation. LTE and 5G RAT have carrier frequencies of 800 MHz and 3.5 GHz. LTE and 5G NR BSs are configured to have transmission power of 40W and 20W. Bandwidth of the LTE and 5G RAT are set to 10 and 20 MHz, respectively. 

\subsection{Simulation results}

\begin{figure}[!t]
\centerline{\includegraphics[width=1\linewidth]{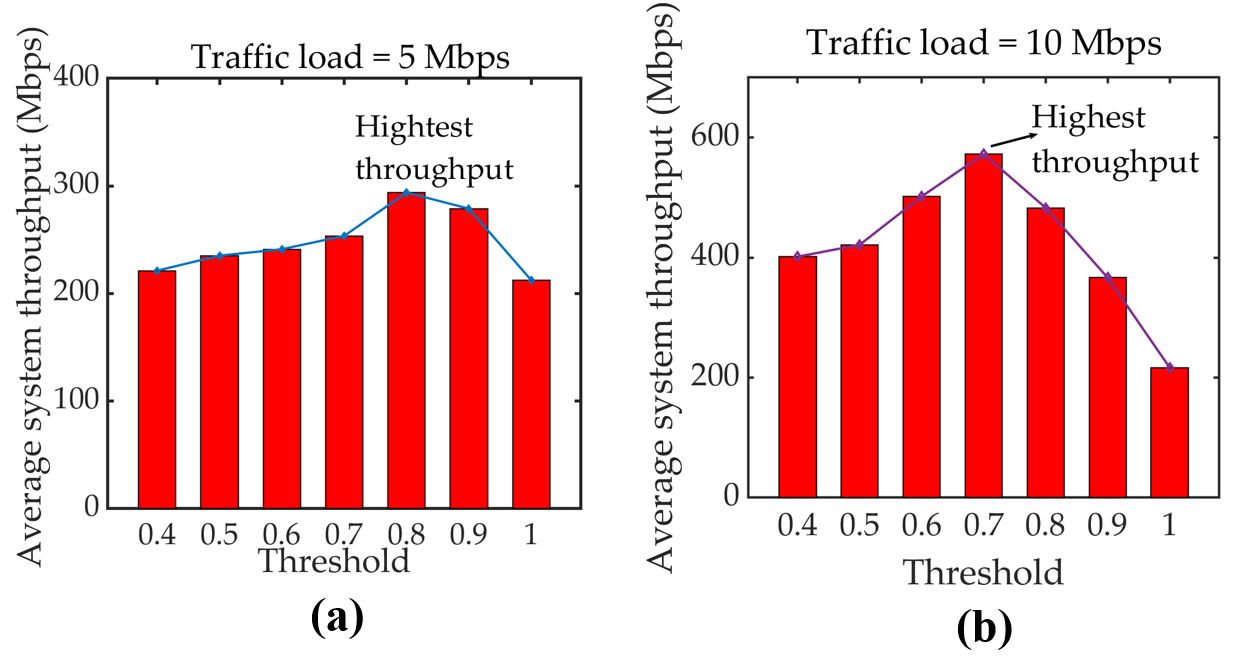}}
\caption{Impact of different thresholds for 5Mbps and 10Mbps traffic load per user.}
\label{fig3}
\vspace{-1.2em}
\end{figure}

Performance evaluation of the proposed HRL-based algorithm is conducted based on three KPIs: packet drop rate, average system throughput, and network delay. The proposed HRL-based method outperforms the threshold-based heuristic and the DRL baseline by gaining 44.57\%  and 24.1\% decrease in the drop rate. Such performance increase by HRL is achieved because of the load-balancing performed using the threshold associated with the queue length along with traffic steering action of the lower level controller. Fig. \ref{fig3} presents us with the impact of different thresholds on average system throughput. We can see that when traffic load is 5Mbps, highest throughput is obtained at 0.8 (80\% of the data queue is occupied). When the threshold is 1, the throughput drastically decreases because the packets are aggregated only when the queue is full and transmission is not possible unless the associated queue is emptied. Similar effects are visible when the traffic load is 10 Mbps except for the fact that the threshold is lower this time (0.7) that gains more output. This is because of the higher arrival rate of the data packets as load increased. 

\begin{figure}[!t]
\centerline{\includegraphics[width=0.7\linewidth]{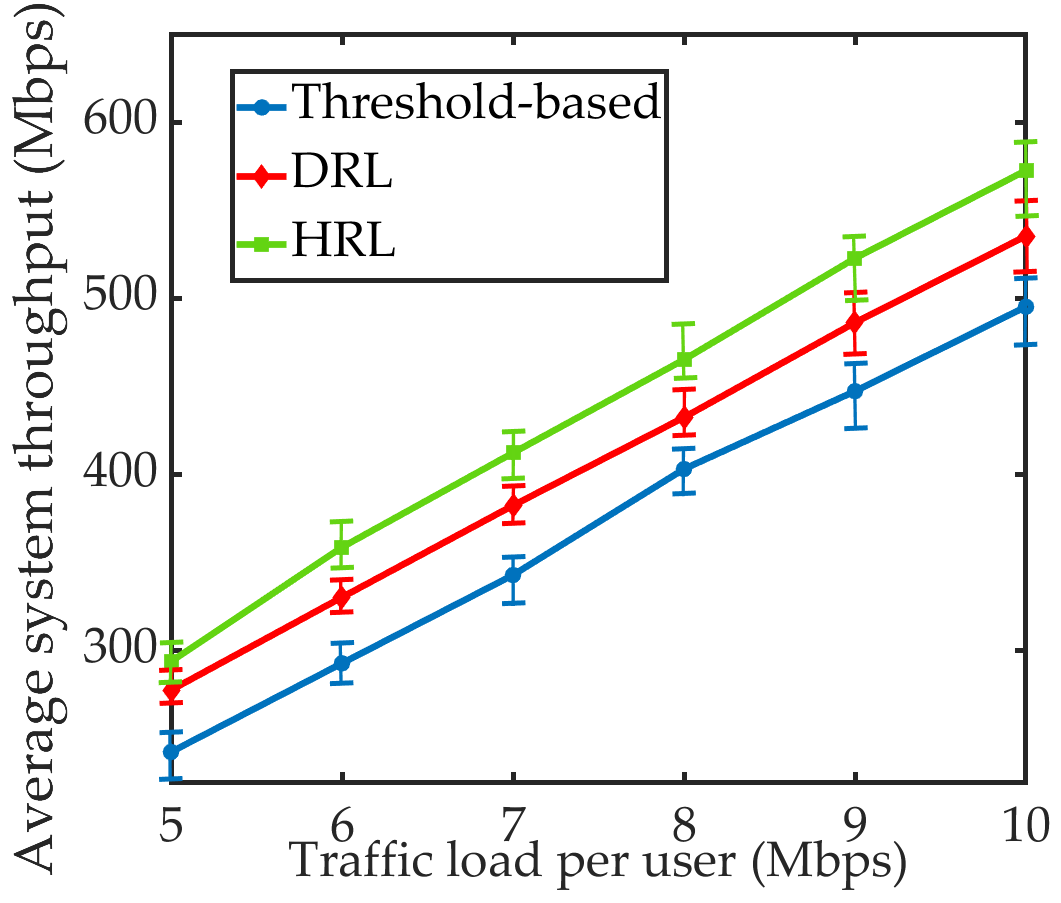}}
\caption{Average system throughput versus traffic load.}
\label{fig4}
\vspace{-1.2em}
\end{figure}

Fig. \ref{fig4} presents a comparative analysis between the proposed HRL-based traffic steering scheme and the baseline algorithms in terms of system throughput. The proposed method outperforms the threshold-based heuristic and the DRL baseline by achieving 12.52\% and 8.49\% increased throughput on an average, respectively. Since DRL is not tailored to handle dynamic change in traffic load  and perform load balancing accordingly, this causes higher packet drop. Hence, this leads to reduction in the overall system throughput. 

\begin{figure}[!t]
\centerline{\includegraphics[width=0.75\linewidth]{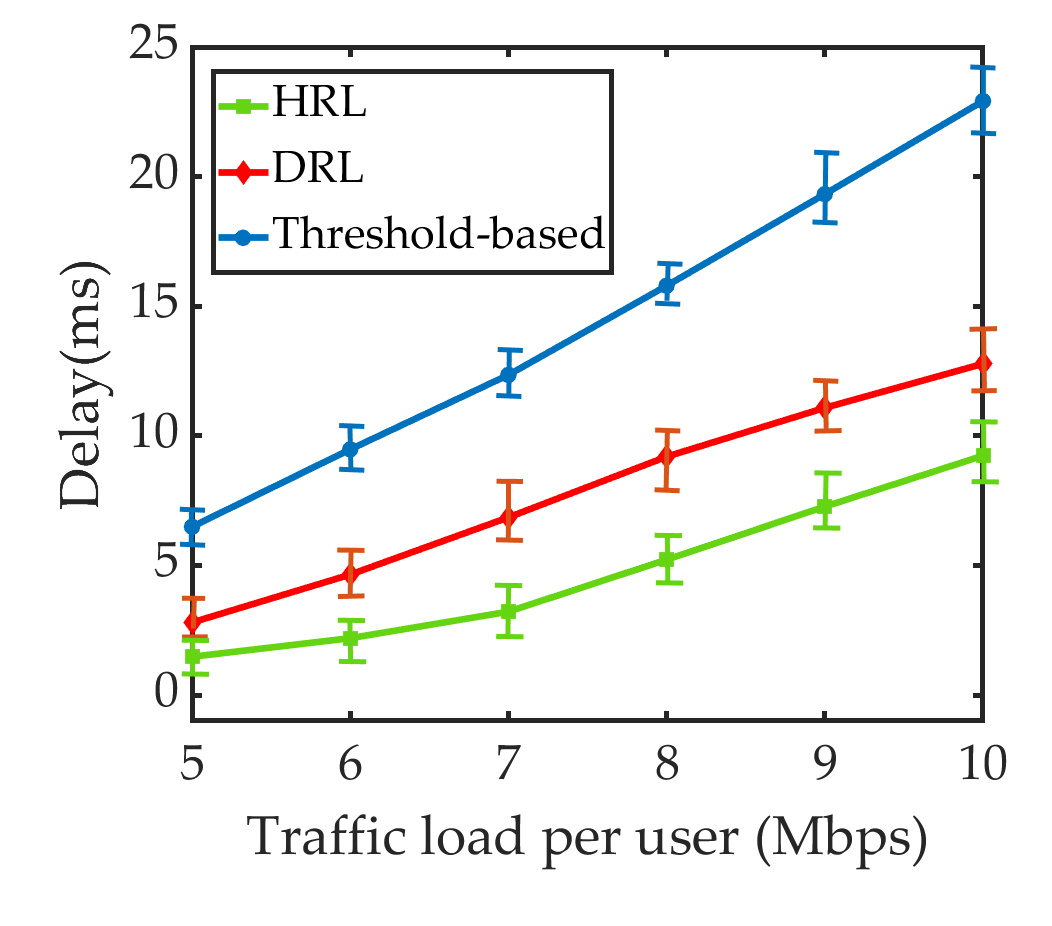}}
\caption{Network delay versus traffic load.} 
\label{fig5}
\vspace{-1em}
\end{figure}

Fig. \ref{fig5} presents the performance comparison among the proposed HRL-based scheme and the baseline algorithms in terms of network delay. The proposed scheme obtains 27.74\% and 39.13\% decrease in network delay compared to the DRL and threshold-based heuristic baseline algorithm, respectively. It is because of the more efficient traffic flow management via threshold-based load balancing at BS level. 

Fig. 6 presents how the traffic is steered to between RATs if the threshold is crossed. High load is imposed when five UEs in 2100th time slot simultaneously gets served by the same base station in the small cell. As a result, fourth UE's data traffic is steered to a different RAT (2450th time slot). Same thing happens at the same time slot for the 6th UE as traffic gets steered to a different RAT.   

\begin{figure}[!t]
\centerline{\includegraphics[width=1\linewidth]{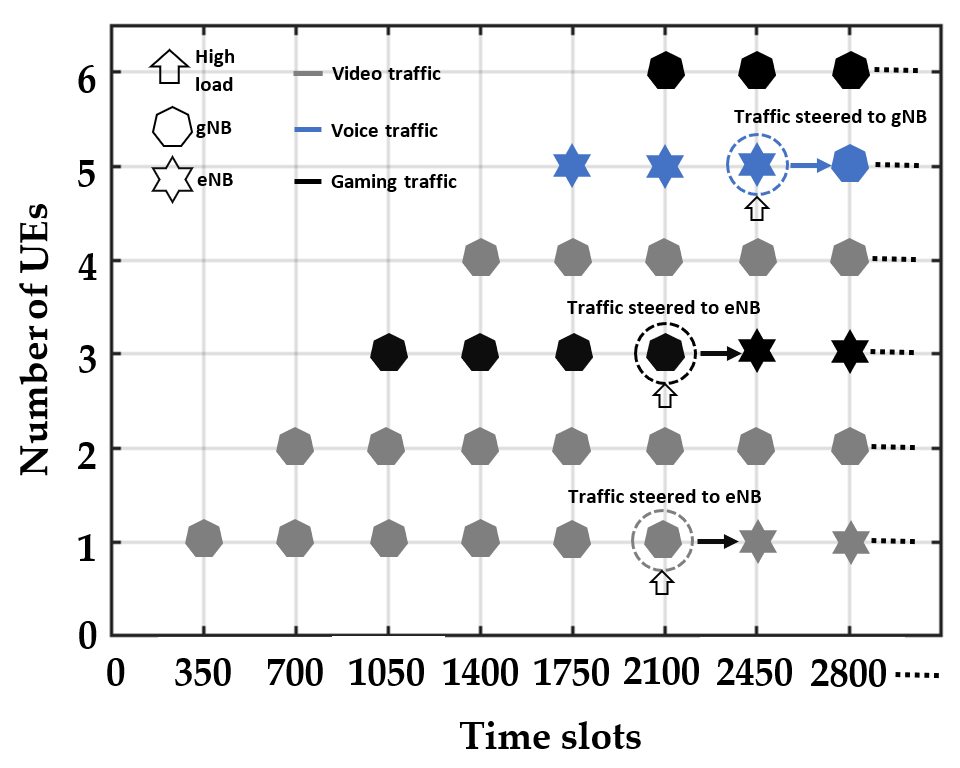}}
\caption{Threshold-based load balancing associated with the queue length.}
\label{fig6}
\vspace{-1.2em}
\end{figure}

\section{Conclusions}
\label{s6}
AI-enabled traffic steering approaches are vastly effective to obtain high performance specially when multi-RAT and multiple traffic types are involved in dense deployments. In this paper, we have proposed a novel load-aware HRL algorithm that can perform QoS-centric, RAT-specific, traffic steering using two levels of controllers, to satisfy the QoS demands of variant traffic types in the network. Optimal threshold selection associated with the queue length of each BS and AI-enabled  traffic steering mechanism has led to 8.49\% (with respect (wrt) to DRL-based baseline), 12.52\% (wrt threshold-based baseline) increase in average system throughput, and 27.74\% (wrt DRL-based baseline), 39.13\% (wrt threshold-based baseline) decrease in network delay. In our future studies, we plan to develop a traffic steering schemes that handle more complex RAN scenarios. 

\section*{Acknowledgement}
This work has been supported by MITACS and Ericsson Canada, and NSERC Canada Research Chairs and NSERC Collaborative Research and Training Experience Program (CREATE) under Grant 497981.

\bibliographystyle{IEEEtran}
\bibliography{ref.bib}
\end{document}